\documentstyle[prl,cite,aps,multicol]{revtex}

\draft
\tighten

\begin{document}

\renewcommand{\narrowtext}{\begin{multicols}{2}
\global\columnwidth20.5pc}
\renewcommand{\widetext}{\end{multicols} \global\columnwidth42.5pc}
\multicolsep = 8pt plus 4pt minus 3pt

\narrowtext

\noindent
{\bf Comment on ``A Convergent Series for the\\
Effective Action of QED''}\\[2ex]

In a recent letter, Cho and Pak claim to have found an 
additional contribution to the quantum electrodynamic one-loop
effective action: a ``logarithmic correction term''
[see the final result in Eq.~(2) and the remark in Ref.~[9]
of \cite{ChPa2001}]. However, the ``logarithmic correction term''
found by Cho and Pak vanishes when the final result is
written in terms of the finite, renormalized, physical 
electron charge. 

Because current determinations of fundamental 
constants~\cite{MoTa2000} rely on renormalized
QED perturbation theory -- without ``logarithmic correction
terms'' of the kind advocated by Cho and Pak --, it is of prime general 
interest to point out that these terms do not appear if on-mass shell
renormalization is used. In the on-mass shell scheme, 
the renormalized QED effective Lagrangian
(see e.g.~\cite[Eq.~(3.43)]{DiGi2000}) reads
\begin{eqnarray}
\label{Leff}
\lefteqn{\small \Delta {\mathcal L} = - \frac{e^2}{8 \pi^2} \,
\lim_{\epsilon,\eta \to 0^{+}} \,
\int_\eta^{{\rm i}\infty + \eta} \,
\frac{{\rm d} s}{s} \,
{\rm e}^{- (m^2 - {\rm i} \epsilon)\, s}} \nonumber \\
&& \small \times
\biggl[  \, a b \, \coth(e a s) \, \cot(e b s) \, -
\frac{a^2 - b^2}{3} - \frac{1}{(e s)^2} \biggr] \,.
\end{eqnarray}
The two latter terms in the integrand are counter\-terms.
The last term simply removes a divergent constant
from the Lagrangian, while the term $-(a^2 - b^2)/3$
-- if it were not removed --, would lead to a logarithmic
divergence at small eigentime $s$. This logarithmically divergent 
term, however,
is proportional to the leading-order Maxwell Lagrangian 
${\mathcal L}_{\mathrm{cl}} = (b^2-a^2)/2$ and leads to a
$Z_3$--renormalization [see Eq.~(8-97) of~\cite{ItZu1980}]. 
Specifically, the introduction of the cut-off 
parameter $\mu$ in~\cite{ChPa2001}
leads to a logarithmic term
$\left[1 - (e^2/12 \pi^2)\,\ln(m^2/\mu^2)\right]$
which multiplies ${\mathcal L}_{\mathrm{cl}}$.
In order to insure compliance with the renormalization
conditions of on-mass shell renormalization 
[see Eqs.~(8-96d) and (8-96e) of~\cite{ItZu1980}],
a further counterterm $+ (e^2/12 \pi^2)\,\ln(m^2/\mu^2) \,
{\mathcal L}_{\mathrm{cl}}$ has to be added to the 
Lagrangian.  As a consequence, the logarithmic correction term is 
absent [see Eqs.\ (2) -- (6) of \cite{ValLamMiel1993}]. 
For the particular problem at hand, the on-mass
shell scheme is well motivated even from a purely mathematical
point of view, as it is evident from the partial fraction theorem
discussed in Sec.~3 of~\cite{JeGiVaLaWe2001}.

If Cho and Pak use different renormalization conditions,
then the logarithmic correction term has to be reabsorbed
into the physical charge of the electron, by considering the 
effect that the term has on matrix elements of transition currents
[see the elucidating discussion on page 325 of~\cite{ItZu1980}].
In this case, we are forced to interpret $e^2_{\rm ph}(\mu) = e^2 \,
[1 - (e^2/12 \pi^2)\,\ln(m^2/\mu^2)] + {\mathcal O}(e^4)$ as the
{\em physical} charge, in which case
the ``logarithmic correction term''~\cite{ChPa2001}
is reabsorbed in a renormalization of charge. 
When expressing $\Delta {\mathcal L}$ in terms of 
$e^2_{\rm ph}(\mu)$ instead of $e^2$, the resulting further
modification of $\Delta {\mathcal L}$ is of the same order as
the two-loop effective Lagragian and 
therefore beyond the validity of the one-loop
approximation inherent to Eq.~(\ref{Leff}).
Finally, we would like to 
remark here that a renormalization-group (RG) improved running of the electron
charge, based on the RG invariance of the 
effective action, has been discussed by Dittrich and Reuter
(Ch.~8 of~\cite{DiRe1985})
and Ritus~\cite{Ri1998}, and that, in the latter case, two-loop
effects are consistently taken into account in the 
analysis of the evolution of the electromagnetic charge.

Finally, we stress that the potentially important 
remark in Ref.~[9] of~\cite{ChPa2001}
falsely suggests that the ``usual'' result 
for $\Delta {\mathcal L}$ given in Eq.~(\ref{Leff}) is incomplete 
without the ``logarithmic correction term''. 
Helpful conversations with H.~Gies, B.~R.~Holstein,
D.~G.~C.~McKeon, C.~Schubert, V.~M.~Shabaev and G.~Soff
are gratefully acknowledged.\\[4ex]
D.~R.~Lamm$^{a)}$, S.~R.~Valluri$^{b)}$,
U.~D.~Jentschura$^{c)}$, and E.~J.~Weniger$^{d)}$\\[2ex]
~$^{a)}${\it Georgia Institute of Technology (GTRI/EOEML), \\
Atlanta, GA 30332-0834} \\
~$^{b)}${\it University of Western Ontario
(Appl. Maths., Physics \& Astronomy), London N6A 3K7, Canada}\\
~$^{c)}${\it TU Dresden, D-01062 Dresden, Germany} \\
~$^{d)}${\it Universit\"{a}t Regensburg (Theoretische Chemie),\\
D-93040 Regensburg, Germany}\\[2ex]

\widetext


\vspace*{-0.7cm}

\begin{thebibliography}{10}
\vspace*{-1.6cm}

\bibitem{ChPa2001}
Y.~M. Cho and D.~G. Pak, Phys. Rev. Lett. {\bf 86},  1947  (2001).

\bibitem{MoTa2000}
P.~J. Mohr and B.~N. Taylor, Rev. Mod. Phys. {\bf 72},  351  (2000).

\bibitem{DiGi2000}
W. Dittrich and H. Gies, {\em Probing the Quantum Vacuum -- Springer
Tracts in Modern Physics Vol. 166} (Springer, Berlin, 2000).

\bibitem{ItZu1980} C. Itzykson and J. B. Zuber, {\em Quantum Field Theory}
(McGraw--Hill, New York, 1980).

\bibitem{ValLamMiel1993}
S.R.\ Valluri, D.R.\ Lamm, and W.J.\ Mielniczuk, Can.\ J.\ Phys.\ {\bf
71}, 389 (1993).

\bibitem{JeGiVaLaWe2001} U. D. Jentschura, H. Gies, S. R. Valluri,
D. R. Lamm and E. J. Weniger, e-print hep-th/0107135, Can. J. Phys.
(2001, in press).

\bibitem{DiRe1985}
W. Dittrich and M. Reuter, {\em Effective Lagrangians in Quantum 
Electrodynamics -- Lecture Notes in Physics Vol. 220}
(Springer, Berlin, 1985).

\bibitem{Ri1998}
V. I. Ritus, {\em Effective Lagrange function of an intense electromagnetic
field in QED}, Proceedings of the conference ``Frontier Tests of QED"",
(Sandansky, Bulgaria, 9--15 June, 1998), Eds. E.~Zavattini, D.~Bakalov,
C.~Rizzo (Heron Press, Sofia, 1998); e-print hep-th/9812124.

\end{thebibliography}
\end{document}